# Ridge, lasso, and elastic-net estimations of the modified Poisson and least-squares regressions for binary outcome data

Takahiro Kitano, DVM, MRes*
The Graduate Institute for Advanced Studies, The Graduate University for Advanced Studies, Tokyo, Japan
ORCID: https://orcid.org/0009-0001-6614-4336

Hisashi Noma, PhD
Department of Interdisciplinary Statistical Mathematics, The Institute of Statistical Mathematics, Tokyo, Japan
The Graduate Institute for Advanced Studies, The Graduate University for Advanced Studies, Tokyo, Japan
ORCID: http://orcid.org/0000-0002-2520-9949

*Corresponding author: Takahiro Kitano
The Graduate Institute for Advanced Studies
The Graduate University for Advanced Studies
10-3 Midori-cho, Tachikawa, Tokyo 190-8562, Japan
TEL: +81-70-2188-9536
e-mail: kitano.takahiro@ism.ac.jp

**Abstract**

Logistic regression is a standard method in multivariate analysis for binary outcome data in epidemiological and clinical studies; however, the resultant odds-ratio estimates fail to provide directly interpretable effect measures. The modified Poisson and least-squares regressions are alternative standard methods that can provide risk-ratio and risk-difference estimates without computational problems. However, the bias and invalid inference problems of these regression analyses under small or sparse data conditions (i.e., the "separation" problem) have been insufficiently investigated. We show that the separation problem can adversely affect the inferences of the modified Poisson and least-squares regressions, and to address these issues, we apply the ridge, lasso, and elastic-net estimating approaches to the two regression methods. As the methods are not founded on the maximum likelihood principle, we propose regularized quasi-likelihood approaches based on the estimating equations for these generalized linear models. The methods provide stable shrinkage estimates of risk ratios and risk differences even under separation conditions, and the lasso and elastic-net approaches enable simultaneous variable selection. We provide a bootstrap method to calculate the confidence intervals on the basis of the regularized quasi-likelihood estimation. The proposed methods are applied to a hematopoietic stem cell transplantation cohort study and the National Child Development Survey. We also provide an R package, `regconfint`, to implement these methods with simple commands.

# 1. Introduction

In clinical and epidemiological studies, logistic regression has been commonly used for binary outcome analyses, and the odds ratios are usually reported (Nurminen 1995). The main reason that logistic regression is adopted is its convenient mathematical properties and computational ease (Agresti 2013). When the event frequencies are rare (<10%), odds ratios can approximate risk ratios, which also justifies the frequent use of odds ratios (Lumley, Kronma, and Ma 2006). However, in the case of non-rare outcomes, odds ratios overestimate risk ratios (Barros and Hirakata 2003) and lack directly interpretable information (Greenland 1987). In such cases, risk ratios and risk differences have been argued to be the preferred epidemiologic effect measures because of their straightforward interpretations and ability to convey clinically important information (Greenland 1987; Sinclair and Bracken 1994; McNutt et al. 2003). Although risk ratios and risk differences can be estimated directly by binomial regression with log and identity links, they suffer from convergence issues and the estimates often cannot be calculated (Wacholder 1986; Mittinty and Lynch 2023). To overcome this problem, researchers have proposed using modified Poisson regression and modified least-squares regression for estimating risk ratios and risk differences, respectively (Zou 2004; Cheung 2007). These methods enable a direct estimation of risk ratios and risk differences using standard statistical software for generalized linear models (GLMs), without the computational issues associated with binomial regression.

Another problem with logistic regression is that it suffers separation and multicollinearity, which can lead to biased estimates and inaccurate variances (Shen and Gao 2008; Goeman, Meijer, and Chaturvedi 2022). Separation often occurs with small or sparse datasets; the problems caused by separation have been referred to as the small or sparse data problem (Greenland, Mansournia, and Altman 2016). Various adjusted estimators have been explored to address this issue, including the ridge, lasso, and elastic-net estimators (Steyerberg,

Eijkemans, and Habbema 2001). In particular, some recent shrinkage estimating methods have been shown to perform effective variable selections simultaneously (Tibshirani 1996; Zou and Hastie 2005). Because the modified Poisson regression and modified least-squares regression are methods used in binomial outcome analyses, they are expected to also suffer from the small or sparse data problem (Albert and Anderson 1984). However, the application of ridge- and lasso-type shrinkage estimations to these two methods has not yet been proposed. A notable issue with these two methods is that their estimates are not ordinary maximum likelihood (ML) estimates; thus, the ordinary ML estimation theory cannot be adapted to these methods.

In this article, we propose using the ridge- and lasso-type shrinkage estimation methods for modified Poisson and modified least-squares regressions. We also provide effective computational methods of risk ratios and risk differences under practical situations for clinical and epidemiological studies and a bootstrapping algorithm to calculate the confidence intervals based on the quasi-ML estimators. We further provide a new R package, *regconfint* (available at https://github.com/kitanota/regconfint), which enables the implementation of these new methods with simple commands. We applied these methods to two real-world examples and investigated whether the estimators improve the accuracy and precision of estimation.

## 2. Methods

*2.1 Modified Poisson and modified least-squares regressions*

Let $y_i$ $(i = 1,2, \dots, n)$ be a binary outcome variable ($y_i = 1$: event occurring, or 0: not occurring) and let $\bm{x}_i = (x_{i1}, x_{i2}, \dots, x_{ip})^T$ $(i = 1,2, \dots, n)$ be $p$ predictor variables for an individual $i$. For the binomial regression models with log and identity links, the probabilities of event occurrence are modeled as

$$\Pr(y_i = 1|\mathbf{x}_i) = \exp(\beta_0 + \beta_1 x_{i1} + \beta_2 x_{i2} + \cdots + \beta_p x_{ip})$$

$$\Pr(y_i = 1|\mathbf{x}_i) = \beta_0 + \beta_1 x_{i1} + \beta_2 x_{i2} + \cdots + \beta_p x_{ip}$$

where $\boldsymbol{\beta} = (\beta_0, \beta_1, \ldots, \beta_p)^T$ are the regression coefficients containing the intercept. The regression coefficients can be estimated by ML estimation within the GLM framework and are interpreted as log risk ratio and risk-difference estimates (Agresti 2013). However, they often suffer from computational difficulties and the ML estimates often cannot be defined. Zou (2004) and Cheung (2007) proposed fitting the ordinary Poisson and least-squares regressions to the binary outcome data and showed that the resultant quasi-ML estimates of the regression coefficients become unbiased estimates for the log risk ratios and risk differences under large samples (Zou 2004; Cheung 2007). These properties are founded on the estimating equation theory of GLM; the quasi-likelihood estimating functions are expressed as

$$U(\boldsymbol{\beta}) = \sum_{i=1}^{n} \boldsymbol{D}_i^T V_i^{-1}(Y_i - \mu_i) = 0 \qquad (*)$$

where $\mu_i$ is the mean function ($= \exp(\boldsymbol{\beta}^T \mathbf{x}_i)$ for the Poisson model and $= \boldsymbol{\beta}^T \mathbf{x}_i$ for the Gaussian model) and $\boldsymbol{D}_i = \partial \mu_i / \partial \boldsymbol{\beta}$; also, $V_i = v(\mu_i)$ is the variance function of the outcome variable ($= \mu_i$ for the Poisson model and $= 1$ for the Gaussian model) ($i = 1, \ldots, n$). The quasi-ML estimator has been shown to become a consistent estimator even if the distributional assumption is failed, as long as the regression function form is correctly specified (Wedderburn 1974; Godambe and Heyde 1987). Note that the variance functions are not correctly specified for the modified Poisson and least-squares regressions; thus, the resultant estimators are not ML estimators or efficient estimators. We therefore refer to them as "quasi-ML" estimators. Only the variance estimators of the regression coefficient estimators should be modified to be the sandwich variance estimator because of the misspecifications of distribution assumptions

(White 1982).

*2.2 Ridge, lasso, and elastic-net estimations*

Shrinkage estimations have been widely used as effective multivariate analysis methods in clinical and epidemiological studies. In particular, some recent shrinkage estimating methods such as the lasso and elastic-net approaches have enabled effective variable selection simultaneously.

*2.2.1 Ridge regression*

Ridge regression was originally proposed to address the problem of collinearity among predictor variables; the ML estimate can be unstable if strong collinearity exists (Schaefer, Roi, and Wolfe 2007). The ridge estimation is a shrinkage estimation performed by the following penalized log likelihood function:

$$\ell(\boldsymbol{\beta}) + \lambda \sum_{j=1}^{p} \beta_j^2$$

where $\lambda\ (>0)$ is a tuning parameter to control the degree of shrinkage. Note that the predictors are usually standardized in advance to have mean 0 and variance 1. The penalty term for ridge regression is the sum of squares of the regression coefficients; the resultant estimate of the regression coefficients is lowered toward zero and can thereby reduce overfitting. Several methods, including cross validation, have been proposed for selecting the tuning parameter $\lambda$ (Hastie, Tibshirani, and Friedman 2009).

*2.2.2 Lasso regression*

Lasso regression is another well-known shrinkage estimation method (Tibshirani 1996). The lasso penalized log likelihood function is defined as

$$\ell(\boldsymbol{\beta}) + \lambda \sum_{j=1}^{p} |\beta_j|$$

where $\lambda$ is a positive tuning parameter similar to that in ridge regression. The penalty term for the lasso function is the sum of the absolute value of the regression coefficients. Note that the predictors are usually standardized in advance to have mean 0 and variance 1. In contrast to ridge regression, lasso regression can shrink some regression coefficients to be exactly 0; therefore, lasso regression can perform shrinkage estimation and variable selection simultaneously (Tibshirani 1996). However, lasso regression is known to perform poorly if highly correlated predictors exist.

*2.2.3 Elastic-net regression*

Elastic-net regression is an effective shrinkage regression method that was proposed to overcome the disadvantage of lasso regression while retaining the variable selection property (Zou and Hastie 2005). The penalty of elastic-net regression is constructed by combining the penalties of lasso and ridge regressions:

$$\ell(\boldsymbol{\beta}) + \lambda \left\{ (1-\alpha) \sum_{j=1}^{p} \beta_j^2 + \alpha \sum_{j=1}^{p} |\beta_j| \right\}$$

where $\lambda\ (>0)$ is the tuning parameter to determine the degree of shrinkage and $\alpha\ (0 \leq \alpha \leq 1)$ is the weight of the lasso and ridge penalties. Even when strong correlations exist among the predictor variables, the strongly correlated predictor variables can remain together for elastic-net regression. Also, similar to lasso regression, some coefficients shrink to exactly 0 when the modified log likelihood function is used (Zou and Hastie 2005). The predictors should also be standardized in advance to have mean 0 and variance 1.

*2.3 Application of shrinkage approaches to the quasi-ML estimation*

Greenland et al. (2016) summarized that four features contribute to the bias caused by small or sparse datasets: few outcome events per variable (EPV), variables with narrow distributions or with categories that are very uncommon, variables that together almost perfectly predict the outcome, and variables that together almost perfectly predict the exposure. The third feature is called separation and is commonly observed in small or sparse datasets. In cases of separation, ML estimates obtained from logistic regression can be seriously biased or even infinite in extreme cases. In addition, confidence intervals derived from such analyses can be uninformative or violate the nominal coverage rates (Heinze 2006).

The regularization approach, which has also been used as a machine learning technique, has been one of the effective data analysis methods used to address these issues (Strickland et al. 2023). In the context of risk prediction, shrinkage estimation methods such as the ridge and lasso approaches have successfully dealt with separation as well (Steyerberg, Eijkemans, and Habbema 2001; Rahman and Sultana 2017).

We here use these shrinkage approaches for quasi-ML estimating methods of the modified Poisson and least-squares regressions. The basic concept is to apply the regularizations to the quasi-likelihood estimating equation (*). Mathematically, these modifications are explained by adding penalization terms to the quasi-log-likelihood functions obtained by integrating the quasi-score function $U(\boldsymbol{\beta})$. The quasi-log-likelihood functions are not uniquely specified because of the ambiguity of the constants of integrations. However, these estimating methods are founded on the ordinary Poisson and least-squares regressions; thus, the most computationally tractable ones are the log-likelihood functions of the Poisson and least-squares

regressions:

$$\ell_1(\boldsymbol{\beta}) = \sum_{i=1}^{n} y_i \log \mu_i - \sum_{i=1}^{n} \mu_i$$

$$\ell_2(\boldsymbol{\beta}) = -\frac{n}{2}\log(2\pi\sigma^2) - \frac{1}{2\sigma^2}\sum_{i=1}^{n}(y_i - \mu_i)^2$$

where $\sigma^2$ is the error variance of the linear regression model. The advantage of using these quasi-log-likelihood functions is that existing statistical software for the regularized estimating methods for GLMs (e.g., *glmnet* (Friedman, Hastie, and Tibshirani 2010)) are all applicable for these regularizations of quasi-likelihood-based estimation. Therefore, we can achieve ridge, lasso, elastic-net, and other effective regularizations for the modified Poisson and least-squares regressions by simply using these procedures for ordinary GLMs.

*2.4 Confidence intervals for the effect measures*

To construct the confidence intervals of the effect measures, bootstrapping methods (Efron 1992; Efron and Tibshirani 1994) can be adopted for all of the shrinkage methods. By adapting nonparametric bootstrap resampling for the cohort data, we can quantify the statistical errors in estimating the regression coefficients involving the selection processes of the regularization parameters (Revan Özkale and Altuner 2022).

The bootstrap confidence intervals are provided using the quantiles of the resultant bootstrap distributions for the regression coefficient estimates; for example, we can use the 2.5th and 97.5th percentiles of them to construct 95% confidence intervals.

*2.5 Software*

To implement these computations involving the bootstrapping approach, we developed the R package *regconfint* (available at https://github.com/kitanota/regconfint) based on the *glmnet* function from the *glmnet* package and the *boot* function from the *boot* package in R. This package allows the regression coefficients of the modified Poisson and least-squares regressions to be estimated via the ridge, lasso, and elastic-net methods. It also involves computational tools for calculating the bootstrap confidence intervals. In addition, we used generic functions that can apply these methods for general GLMs with various link functions.

## 3. Applications

To illustrate their effectiveness, we applied the proposed shrinkage estimating methods of the modified Poisson and least-squares regressions to two cohort datasets with different data sizes and whose event frequencies were not small: the Retrospective Cohort Study of the Effects of Donor killer immunoglobulin-like receptors (KIR) genotype on the reactivation of cytomegalovirus (CMV) after myeloablative allogeneic hematopoietic stem cell transplant (Sobecks et al. 2011), and the National Child Development Survey (NCDS) (Power and Elliott 2006).

*3.1 Retrospective cohort study for hematopoietic stem cell transplantation*

As a real-world example with non-rare outcomes and small data size, we considered a retrospective cohort study that investigated whether donor KIR genotype influenced the reactivation of CMV after T-cell replete, matched sibling donor reduced-intensity conditioning allogeneic hematopoietic stem cell transplantation (HSCT) (Sobecks et al. 2011; Higgins 2022). In this cohort dataset for a total 64 patients, the number of events and the event frequency were 26 and 40.6%, respectively. Following the original study of Sobecks et al. (2011), we used the occurrence of CMV reactivation posttransplant as the binary outcome and carried out modified Poisson and least-squares regression analyses that modeled 10 explanatory variables. In the original study, the primary risk factor of interest was the number of activating killer immunoglobulin-like receptors (aKIRs, 1–4 vs 5–6). Of the 10 variables, one variable (Age) was continuous, one variable (Diagnosis of cancer) was categorical, and the remaining 8 variables were binary. Three variables (Type of cancer diagnosed, Total body irradiation dosage, and aKIRs) were turned into dummy variables, and a total of 21 explanatory variables were then included in the regression models. The variables in the CMV dataset are summarized in Supplementary Table 1. Separations were observed in four variables among the types of cancer diagnosed (Acute lymphoblastic leukemia, Aplastic anemia, Congenital anemia, and Myeloproliferative disorder). The variance inflation factor (VIF) was less than 10 for all of the variables, indicating the absence of strong multicollinearity; the EPV was 1.24 (26/21).

First, we applied conventional binomial regression models with log-link and identity-link functions; however, the ML estimates were not obtained because of the convergence problem. We then adapted the modified Poisson regression and least-squares regressions with the ordinary quasi-ML and shrinkage estimations (i.e., ridge, lasso, and elastic-net estimations). We also applied the logistic regression analysis for comparative purposes (i.e., to compare the resultant odds-ratio estimates with the risk-ratio estimates). The shrinkage estimation was conducted using the *glmnet* package of R (Friedman, Hastie, and Tibshirani 2010; Tay, Narasimhan, and Hastie 2023); the regularization parameters were estimated by 10-fold cross validations. All of the explanatory variables were standardized to have mean 0 and variance 1 when the model was fit, and the resultant estimates were returned on the original scales for reports. The bootstrap 95% confidence intervals of the shrinkage estimation were calculated using the *boot* package of R with 1,000 resamplings (Canty and Ripley 2022). The illustrative R code used in this study is available at https://github.com/kitanota/regconfint.

Figure 1 presents forest plots of odds-ratio estimates from logistic regression, and risk-ratio estimates from modified Poisson regression with quasi-ML estimation and shrinkage estimation. As anticipated, the magnitudes of associations were markedly large in the odds ratios compared with those in the risk ratios; the odds-ratio estimates could overestimate the risk ratios, making logistic regression analysis unsuitable for these applications. For the variables with separation, odds-ratio estimates and risk-ratio estimates from the ordinary ML or quasi-ML estimation were not obtained. In addition, for many variables, the 95% confidence intervals were too wide. For example, the risk-ratio estimate and its 95% confidence interval of aKIRs, the primary factor, was 0.38 (0.10–1.43). However, the ridge, lasso, and elastic-net methods provided interpretable risk-ratio estimates and much narrower confidence intervals. The point estimates of risk ratios obtained from those shrinkage estimation methods were clearly shrunken from those obtained through the conventional quasi-ML estimates in the directions of null associations. The risk ratio and its 95% confidence interval of aKIRs were estimated as 0.80 (0.25–1.00) with the ridge method. Through lasso and elastic-net estimations, aKIRs and three other variables (Sex, HLW-Cw group, and Total body irradiation dosage) remained for the variable selection processes; the point estimates and 95% confidence intervals of aKIRs were 0.90 (0.13–1.00) and 0.90 (0.17–1.00), respectively. The weight of the lasso and ridge penalties ($\alpha$) of the elastic-net estimation was 0.93, indicating that the performance of the elastic-net method was similar to that of the lasso method.

Figure 2 shows forest plots of the risk-difference estimates from the modified least-squares regression with quasi-ML estimation and with shrinkage estimations. Similar to the risk-ratio estimates, the risk-difference estimates from the ridge, lasso, and elastic-net regressions were markedly reduced toward 0 and the shrinkage estimations provided narrower confidence intervals compared with the quasi-ML estimation. The ridge estimate of the risk difference and its confidence interval of aKIRs was −0.09 (−0.38–0.00). The lasso and elastic-net methods selected 11 variables, including aKIRs, and the risk-difference estimates and 95% confidence intervals of the aKIRs were −0.16 (−0.46–0.00) and −0.15 (−0.44–0.00), respectively. Interestingly, the estimated $\alpha$ was 0.24, indicating that the elastic-net method performed more like the ridge method even though the results of the two regressions were similar. Overall, the proposed method effectively provided interpretable estimates of risk ratio and risk difference and stable interval estimates, even under the setting of a small dataset with non-rare outcomes and with the presence of separation.

*3.2 National Child Development Survey*

As an example of a study with non-rare outcomes and a large sample size, we considered the applications to the NCDS. The NCDS is a cohort study that started as a study of perinatal mortality, focusing on the births in a single week of 1958 in England, Scotland, and Wales (thus also referred as the 1958 birth cohort), and evolved into a longitudinal study to monitor participants' educational, physical, and social development (Power and Elliott 2006). We adopted the dataset derived from the study on the relationship between educational qualifications in the United Kingdom and wage returns (Battistin and Sianesi 2011; Zhou et al. 2022). We adopted a binary outcome of wage—the dichotomized wage with the cutoff of the hourly wage of actively employed British males aged 30–39 in 1991—and included 14 explanatory variables in the analysis. The number of events and the event frequency in the NCDS datasets were 1,610 and 44.2%, respectively, with a sample size of 3,642. Of the 14 variables, 9 variables were continuous (Years of education(father), Years of education(mother), Age of father in 1974, Age of mother in 1974, Number of siblings, Math test score at age 7, Math test score at age 11, Reading test score at age 7, and Reading test score at age 11), 2 variables (Academic qualification and School type at age 16) were categorical, and 3 variables (Race, Any academic qualification, and Employment of mother in 1974) were binary. For the two categorical variables, we generated dummy variables and used them for the analysis. In total, 18 explanatory variables were included in the subsequent analyses. The variables in the NCDS dataset are summarized in Supplementary Table 2. For this dataset, separation was not observed; the VIF was less than 10 for all variables, which indicated the absence of strong multicollinearity, and the EPV was 89.4 (1610/18).

We first applied the binomial regressions with log-link and identity-link functions for the NCDS dataset. However, even for the large dataset, the regressions failed to converge and the ML estimates were not obtained. We then conducted the same analyses as those applied to the CMV dataset: the modified Poisson regression and least-squares regression with the ordinary quasi-ML and shrinkage estimations (ridge, lasso, and elastic-net estimations) and the conventional logistic regression analysis. For computations, we also used the *glmnet* package (Friedman, Hastie, and Tibshirani 2010; Tay, Narasimhan, and Hastie 2023) and the *boot* package (Canty and Ripley 2022). The bootstrap 95% confidence intervals of the regression coefficients were computed by 1,000 resamplings.

Figure 3 presents forest plots of odds-ratio estimates from logistic regression and the risk-ratio estimates from the modified Poisson regression with conventional ML and quasi-ML estimation and with shrinkage estimations. In this case, the odds-ratio estimates were also remarkably large compared with the risk-ratio estimates, and the logistic regression is inappropriate for non-rare event cases. Also, the risk-ratio estimates were generally shrunken to the null associations (= 1.0) by the shrinkage estimations and the 95% confidence intervals estimated by the shrinkage estimations were narrower than those obtained by the quasi-ML estimation. In particular, for the "Academic qualification (Advanced)" variable, which had the strongest association with the outcome, the risk-ratio estimates were clearly shrunken toward the null by the shrinkage estimations: quasi-ML estimation 1.81 (1.47–2.23); ridge 1.67 (1.44–1.92); lasso 1.63 (1.48–2.12); elastic-net 1.62 (1.48–2.10). In the lasso and elastic-net estimations, 12 variables were finally selected, including "Academic qualification (Advanced)." The optimal $\alpha$ of the elastic-net estimations, as obtained by cross-validation, was 0.58, indicating that it performed more like the lasso estimation.

Figure 4 illustrates the forest plots of risk differences estimated from the modified least-squares regressions using quasi-ML estimation and shrinkage estimations. While the degrees of shrinkage were not remarkable compared with the risk-ratio cases, the risk-difference estimates were certainly shrunken to null and the confidence intervals were generally narrower for the shrinkage estimations. For example, the risk-difference estimates and their 95% confidence intervals for "Academic qualification (Advanced)" were 0.17 (0.12–0.23) for quasi-ML estimation; 0.16 (0.12–0.21) for ridge; 0.18 (0.12–0.23) for lasso; 0.18 (0.12–0.23) for elastic-net. The hyperparameter α of the elastic-net estimation was specified as 0.58 by cross-validation, indicating that it performed more like the lasso estimation. For the lasso and elastic-net estimations, the tuning parameter was 0.00046 and 0.00087, respectively, and no variables were dropped for either estimation. This result suggests that all the predictors had certain associations with the outcome variable in the estimation of risk difference with this example. In summary, the proposed method successfully provided interpretable estimates of risk ratio and risk difference and stable interval estimates with moderate data size and non-rare outcomes.

## 4. Discussion

In many clinical and epidemiological studies, binary outcomes are commonly adopted; however, the event frequencies are often not rare (> 10%). In such situations, odds-ratio estimates by logistic regression usually overestimate the risk ratios and fail to provide interpretable epidemiological information (Greenland 1987; Barros and Hirakata 2003). Even for randomized trials, the logistic regression might provide uninterpretable effect measure estimates. The modified Poisson and least-squares regressions are effective alternative methods that provide directly interpretable effect measure estimates. In this study, we proposed regularized regression methods (i.e., the ridge, lasso, and elastic-net methods) for these quasi-likelihood-based regression analyses. We found that, like the ML estimation of logistic regression, the ordinary quasi-ML estimations have numerical instability and possible biases under small-sample or separation conditions. By contrast, the proposed methods provide stable and adequately shrunken estimates under these situations, as expected. The confidence intervals were also suitably estimated, indicating that the accuracy and precision could be improved by the proposed methods. There are several possible situations in which epidemiological research needs to be conducted with a limited number of patients (e.g., rare diseases, emerging infectious diseases, and the early stages of pandemics) (Norrie 2020; Raycheva et al. 2023; Wang et al. 2023), and the proposed methods will provide useful solutions for these practical situations.

The convergence rates for exploring the solutions of estimating equations were also generally good because the original estimating functions are the well-known GLM's quasi-likelihood functions with canonical links. Although the conventional binomial log-linear and linear models could not provide ML estimates for both of the example datasets, all of the proposed methods could provide regularized risk-ratio and risk-difference estimates. These favorable characteristics are a general property of our methods.

The ease of computation is also a remarkable advantage of the proposed methods, which can be carried out using existing statistical packages for shrinkage methods for GLMs (e.g., *glmnet* in R). In addition, we developed a new R package, *regconfint*, that allows for the estimation of bootstrap confidence intervals with a simple command. A potential difficulty is a large computational burden of the elastic-net regression that involves the selecting process of $\alpha$ by cross-validation, and the uncertainty of $\alpha$ should be considered in calculating bootstrap confidence intervals. However, we introduced parallel computation procedures in the bootstrap replication procedure in the *regconfint* package to address this issue, enabling a substantial reduction in computation time. These computational tools should facilitate the use of these advanced statistical techniques for practitioners in clinical and epidemiological studies involving non-statisticians.

A possible topic for future studies is to investigate similar regularized estimating methods for other effective approaches recently proposed to improve convergences in binomial log-linear models for estimating risk ratios (Mittinty and Lynch 2023; Zhu et al. 2023). Also, several other effective extensions might be realized for the binomial log-linear and linear models. In any event, the overall principles of the proposed methods provided in this work can be generically applied to various estimating-equation-based approaches.

## 5. Conclusion

The ridge, lasso, and elastic-net shrinkage estimating methods applied to the modified Poisson and modified least-squares regressions can be a valuable option in multivariate analyses for clinical and epidemiological studies to address problems with small or sparse data. These methods can be easily implemented using a newly developed R package, *regconfint*, with simple computational commands.

Funding

This study was supported by Grants-in-Aid for Scientific Research from the Japan Society for the Promotion of Science (grant numbers: JP23K11931, JP22H03554, and JP23H03063).

Figure 1 Forest plot of odds ratios and risk ratios estimated with the CMV dataset

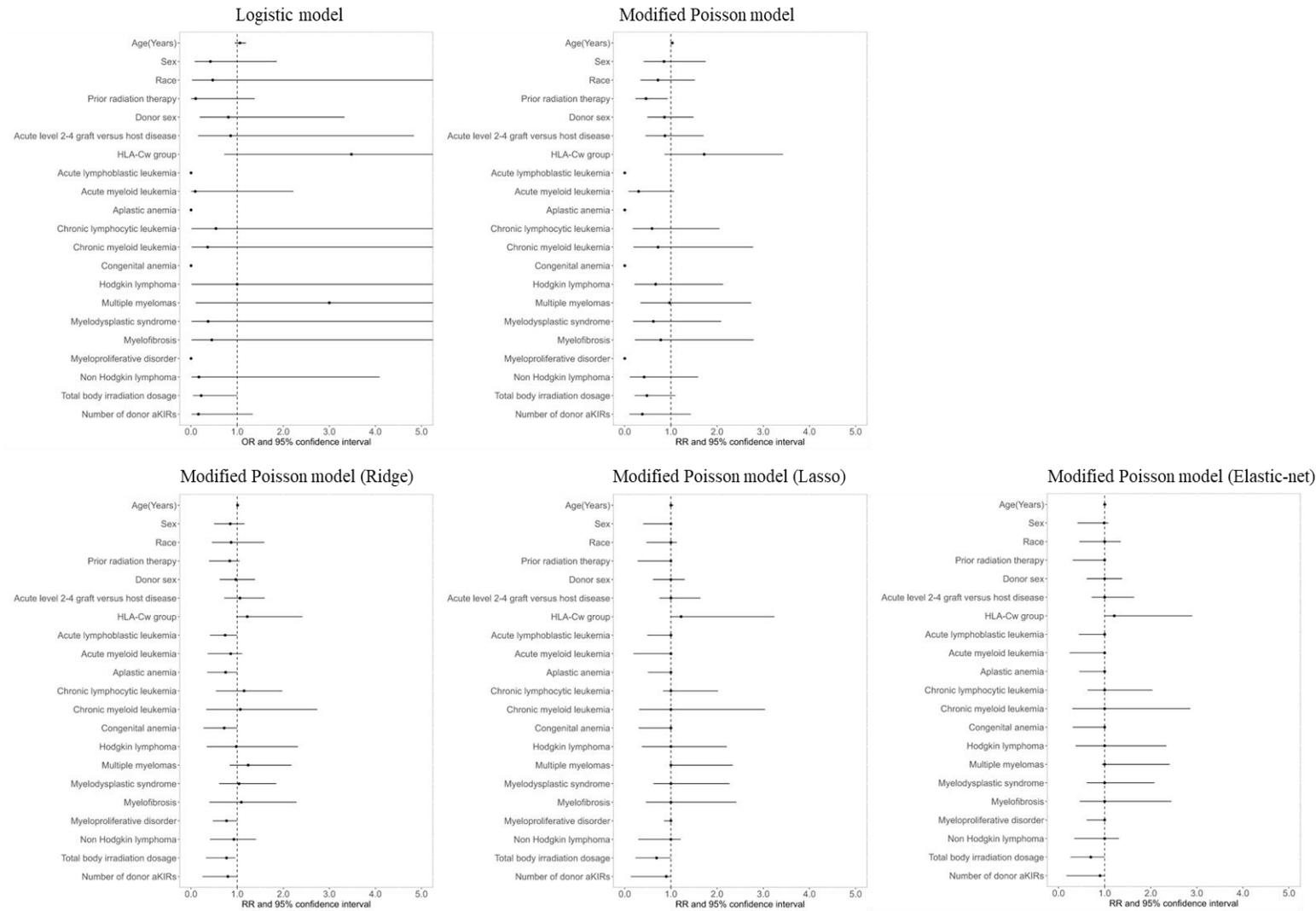

Figure 2 Forest plot of risk differences estimated with the CMV dataset

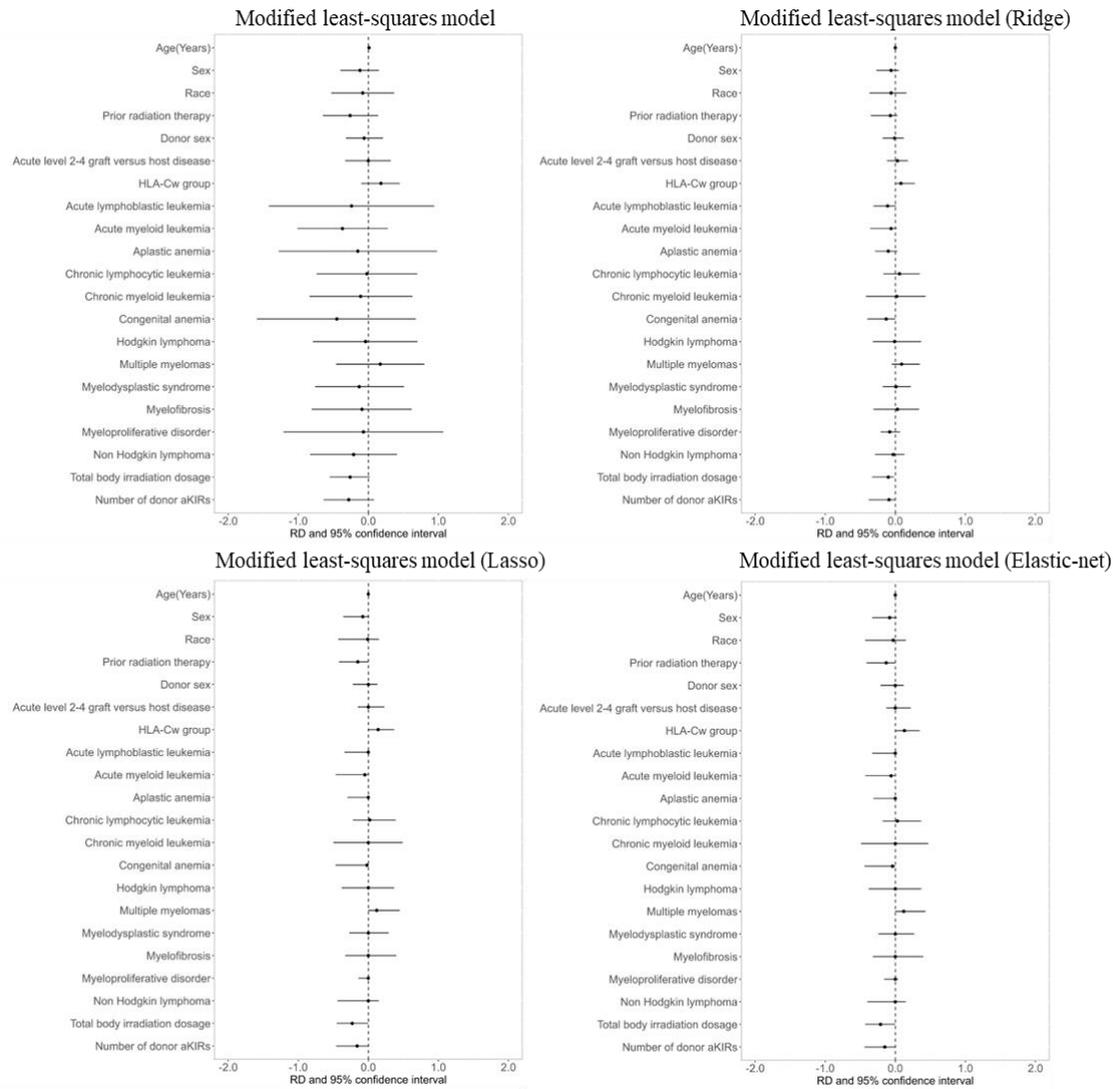

Figure 3 Forest plot of odds ratios and risk ratios estimated with the NCDS dataset

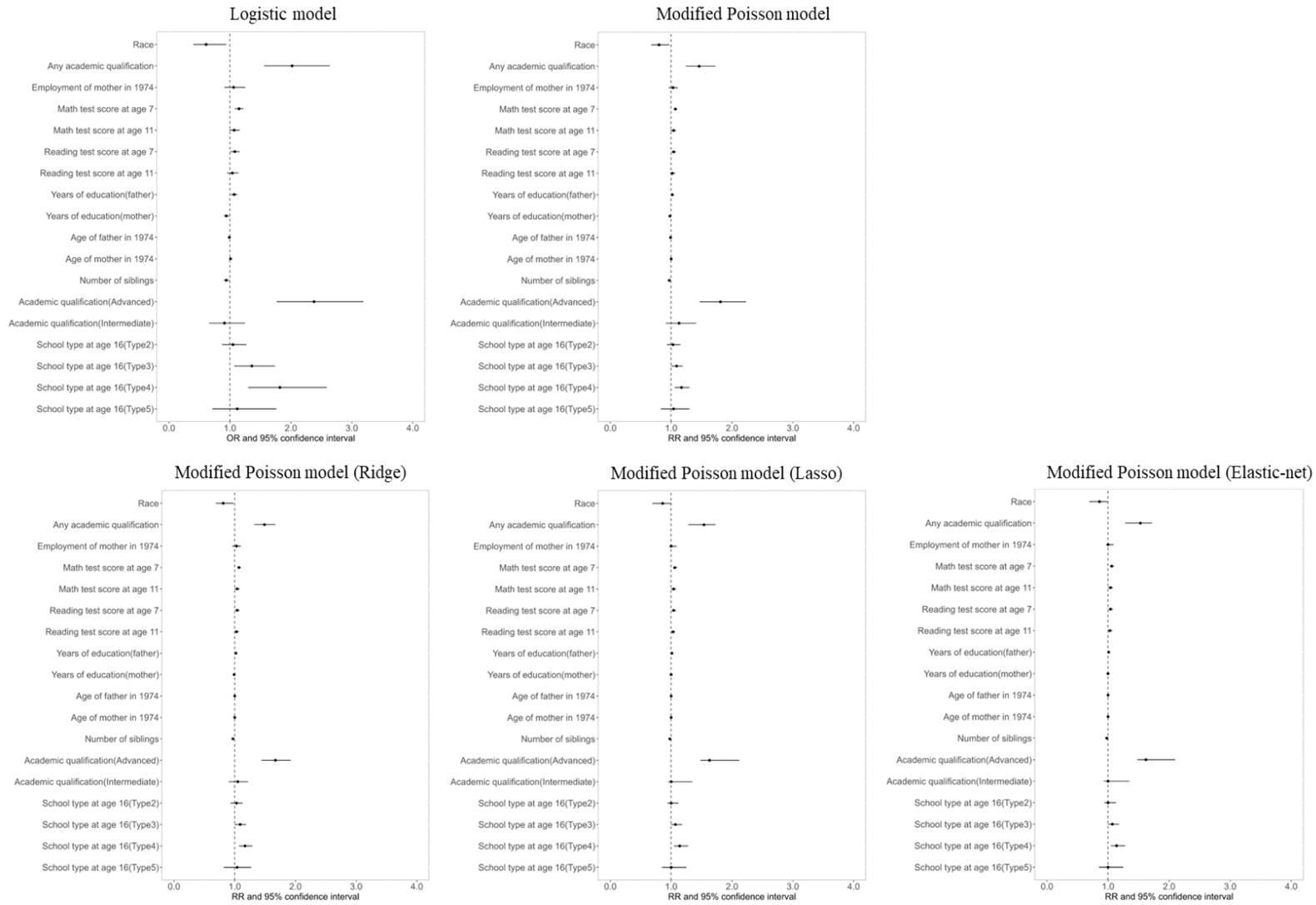

Figure 4 Forest plot of risk differences estimated with the NCDS dataset

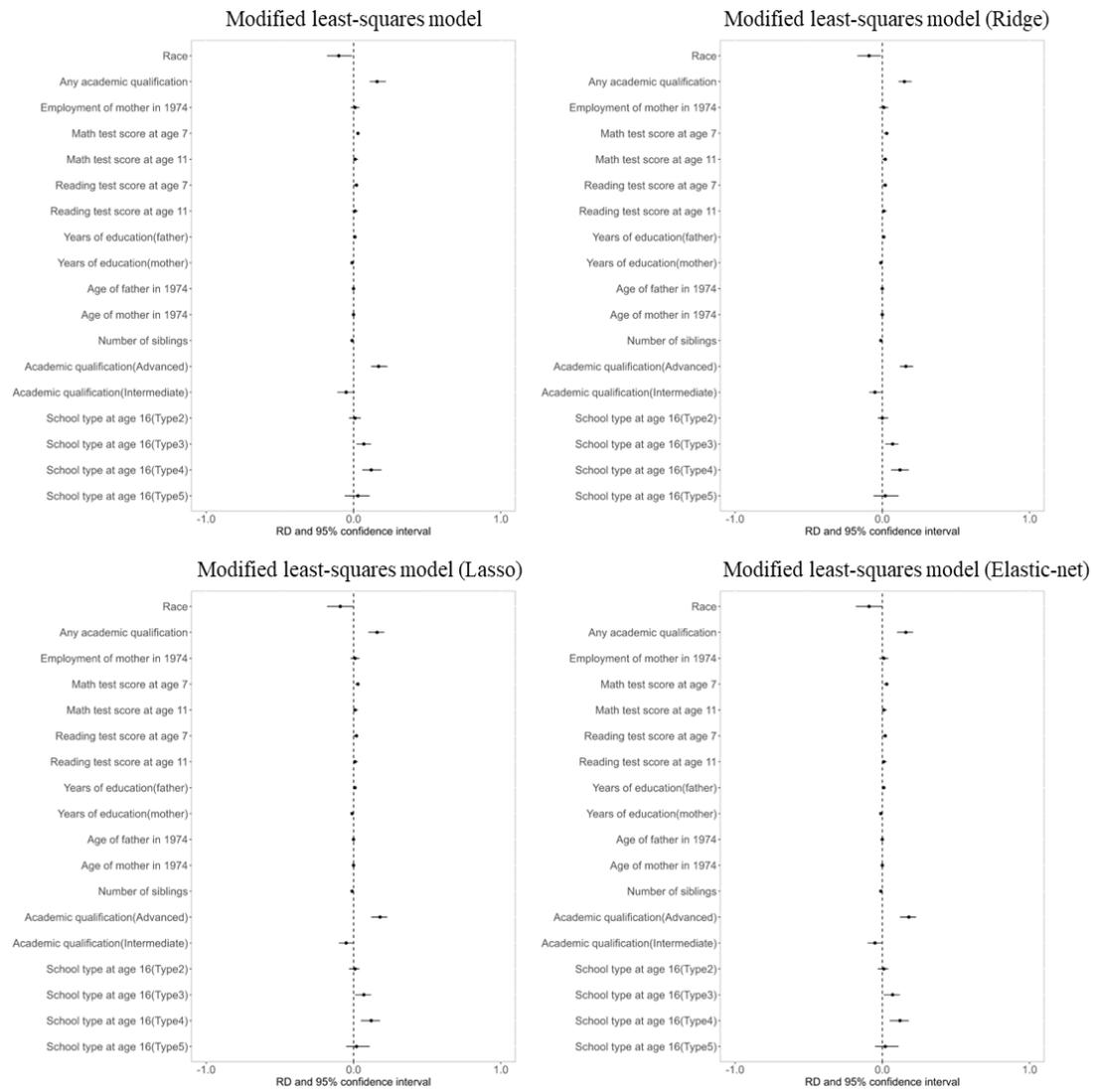

Supplementary Table 1 Characteristics of the CMV dataset

| Characteristic | Infection of cytomegalovirus | |
|---|---|---|
| | Yes, N = 26[1] | No, N = 38[1] |
| Age (years) | 53.3 (8.2) | 51.8 (10.2) |
| Sex | | |
|   Female | 16 / 26 (61.5%) | 14 / 38 (36.8%) |
|   Male | 10 / 26 (38.5%) | 24 / 38 (63.2%) |
| Race | | |
|   African-american | 4 / 26 (15.4%) | 2 / 38 (5.3%) |
|   White | 22 / 26 (84.6%) | 36 / 38 (94.7%) |
| Type of cancer diagnosed | | |
|   Acute lymphoblastic leukemia | 0 / 26 (0.0%) | 1 / 38 (2.6%) |
|   Acute myeloid leukemia | 3 / 26 (11.5%) | 9 / 38 (23.7%) |
|   Aplastic anemia | 0 / 26 (0.0%) | 1 / 38 (2.6%) |
|   Chronic lymphocytic leukemia | 3 / 26 (11.5%) | 2 / 38 (5.3%) |
|   Chronic myeloid leukemia | 2 / 26 (7.7%) | 2 / 38 (5.3%) |
|   Congenital anemia | 0 / 26 (0.0%) | 1 / 38 (2.6%) |
|   Hodgkin lymphoma | 1 / 26 (3.8%) | 2 / 38 (5.3%) |
|   Multiple myelomas | 5 / 26 (19.2%) | 2 / 38 (5.3%) |
|   Myelodysplastic syndrome | 4 / 26 (15.4%) | 5 / 38 (13.2%) |
|   Myelofibrosis | 2 / 26 (7.7%) | 2 / 38 (5.3%) |
|   Myeloproliferative disorder | 0 / 26 (0.0%) | 1 / 38 (2.6%) |

|  | Infection of cytomegalovirus | |
|---|---|---|
| **Characteristic** | **Yes**, N = 26[1] | **No**, N = 38[1] |
| Non-hodgkin lymphoma | 4 / 26 (15.4%) | 8 / 38 (21.1%) |
| Renal cell carcinoma | 2 / 26 (7.7%) | 2 / 38 (5.3%) |
| Prior radiation therapy | | |
| No | 23 / 26 (88.5%) | 30 / 38 (78.9%) |
| Yes | 3 / 26 (11.5%) | 8 / 38 (21.1%) |
| Donor sex | | |
| Female | 13 / 26 (50.0%) | 19 / 38 (50.0%) |
| Male | 13 / 26 (50.0%) | 19 / 38 (50.0%) |
| Total body irradiation dosage | | |
| 200 | 20 / 26 (76.9%) | 16 / 38 (42.1%) |
| 400 | 6 / 26 (23.1%) | 22 / 38 (57.9%) |
| Number of donor activating killer immunoglobulin-like receptors | | |
| 1-4 | 23 / 26 (88.5%) | 25 / 38 (65.8%) |
| 5-6 | 3 / 26 (11.5%) | 13 / 38 (34.2%) |
| Acute level 2-4 graft versus host disease | | |
| No | 14 / 26 (53.8%) | 24 / 38 (63.2%) |
| Yes | 12 / 26 | 14 / 38 |

|                      | Infection of cytomegalovirus |                |
|----------------------|:---------------------------:|:--------------:|
| **Characteristic**   | **Yes**, N = 26[1]          | **No**, N = 38[1] |
|                      | (46.2%)                     | (36.8%)        |
| HLA-Cw group         |                             |                |
|   Heterozygous | 8 / 26 (30.8%)          | 23 / 38 (60.5%) |
|   Homozygous   | 18 / 26 (69.2%)         | 15 / 38 (39.5%) |

[1]Mean (SD); n / N (%)

Supplementary Table 2 Characteristics of the NCDS dataset

| Characteristic | Wage | |
|---|---|---|
| | **Above average**, N = 1,610[1] | **Below average**, N = 2,032[1] |
| Race | | |
|   Other | 53 / 1,610 (3.3%) | 60 / 2,032 (3.0%) |
|   White | 1,557 / 1,610 (96.7%) | 1,972 / 2,032 (97.0%) |
| Any academic qualification | | |
|   No | 274 / 1,610 (17.0%) | 969 / 2,032 (47.7%) |
|   Yes | 1,336 / 1,610 (83.0%) | 1,063 / 2,032 (52.3%) |
| Academic qualification | | |
|   Advanced | 1,121 / 1,610 (69.6%) | 685 / 2,032 (33.7%) |
|   Intermediate | 165 / 1,610 (10.2%) | 730 / 2,032 (35.9%) |
|   No | 324 / 1,610 (20.1%) | 617 / 2,032 (30.4%) |
| Employment of mother in 1974 | | |
|   No | 757 / 1,610 (47.0%) | 1,017 / 2,032 (50.0%) |
|   Yes | 853 / 1,610 (53.0%) | 1,015 / 2,032 (50.0%) |
| School type at age 16 | | |
|   Type1 | 859 / 1,610 (53.4%) | 1,288 / 2,032 (63.4%) |
|   Type2 | 260 / 1,610 (16.1%) | 476 / 2,032 (23.4%) |
|   Type3 | 292 / 1,610 (18.1%) | 148 / 2,032 (7.3%) |
|   Type4 | 158 / 1,610 (9.8%) | 57 / 2,032 (2.8%) |
|   Type5 | 41 / 1,610 (2.5%) | 63 / 2,032 (3.1%) |
| Math test score at age 7 | 3.6 (1.3) | 2.9 (1.4) |
| Math test score at age 11 | 3.7 (1.3) | 2.9 (1.4) |
| Reading test score at age 7 | 3.4 (1.3) | 2.7 (1.3) |

|  | Wage | |
| --- | --- | --- |
| **Characteristic** | **Above average**, N = 1,610[1] | **Below average**, N = 2,032[1] |
| Reading test score at age 11 | 3.6 (1.3) | 2.7 (1.4) |
| Years of education(father) | 9.2 (4.1) | 8.5 (3.5) |
| Years of education(mother) | 9.0 (3.7) | 8.5 (3.4) |
| Age of farther in 1974 | 46.5 (6.1) | 46.7 (6.6) |
| Age of mother in 1974 | 43.6 (5.4) | 43.6 (5.7) |
| Number of siblings | 1.5 (1.6) | 1.9 (1.9) |

[1] n / N (%); Mean (SD)